# Nature of photoexcited states in ZnO-embedded graphene quantum dots

Ivan Shtepliuk* and Rositsa Yakimova

*Semiconductor Materials Division, Department of Physics, Chemistry and Biology-IFM, Linköping University, S-58183 Linköping, Sweden*

**Abstract**

The combination of wide-band gap semiconductors like zinc oxide (ZnO) and graphene quantum dots (GQDs) is a promising strategy to tune optoelectronic properties of GQDs and to develop new functionalities. Here we report on a theoretical design of not-yet-synthesized hybrid materials composed of ZnO clusters surrounded by carbon moieties, hereinafter referred to as ZnO-embedded graphene quantum dots. Their structure and light absorption properties are presented, with an in-depth analysis of the nature of the photoexcited states. The stability of the $(ZnO)_nC_{96-2n}$ system with $n$=1, 3, 4, 7, 12 and 27 is investigated by estimating cohesive energy and performing vibrational mode analysis. A strong dependence of the structural and optoelectronic properties of the hybrid material on the amount of ZnO pairs is revealed and discussed. A strong light absorption and unexpected enhancement of Raman modes related to the vibrations in carbon moiety are observed for highly symmetric $(ZnO)_{27}C_{42}$ system that makes it an ideal study subject. Complementary excited state analysis, charge density difference (CDD) analysis and interfragment charge transfer analysis enabled reaching deep insights into the nature of the excited states. A dominating contribution of doubly degenerate locally excited states in broadband light absorption by $(ZnO)_{27}C_{42}$ is identified. The present results are helpful to elucidate the nature of the fundamental internal mechanisms underlying the light absorption in ZnO-embedded graphene quantum dots, thereby providing a scientific background for future experimental study of low-dimensional metal-oxygen-carbon materials family.

*Corresponding author: ivan.shtepliuk@liu.se

# Introduction

Graphene quantum dots (GQDs) – nanosized graphene fragments – are currently considered as an exclusive organic material that brings together the beneficial properties of graphene (large surface area, high carrier mobility and conductivity) and the excellent light-emitting ability dictated by its semiconducting character. Possessing tuneable band gap, low toxicity, and good chemical stability, GQDs find applications in biomedicine, optoelectronics, energy, and sensing technologies. A further integration of zero-dimensional (0D) GQDs with inorganic compound semiconductors (ICS) is a winning strategy to create advanced organic-inorganic hybrid nanomaterials with novel functionalities. Among different ICS, zinc oxide (ZnO) is of special importance for many applications due its unique physical properties (direct bandgap of 3.3 eV at room temperature; a strong radiation hardness; large exciton binding energy of 60 meV; antimicrobial activity etc.), relative growth/synthesis simplicity, and morphological diversity[1,2].

Over the recent years a great deal of progress has been made in synthesis, material characterization and conceptualization of ZnO-GQDs nanohybrids. More specifically, a combination of both materials enables to design high-performance sensors for detection of different substances including $H_2O_2$[3], $CO_2$[4], ammonia[5], ethanol[6], hydroquinone[7], zearalenone in mildewing cereal crops[8], acetone[9,10], $H_2S$[11], acetic acid[12], chemo-therapeutic agent (6-mercaptopurine)[13], and $NO_2$[14,15]. Leaving behind brackets various sensing principles that have been applied for quantification of gas or liquid molecules, it should be emphasized that in most cases sensitive hybrid materials were constructed through functionalization or decoration of host ZnO matrix with undoped or doped GQDs. Another example of a beneficial synergy between GQDs and ZnO features concerns a high promise of related nanohybrids for design of next-generation ultraviolet (UV) photodetectors[16–29]. On one hand, the presence of GQDs may favour separation and transport of the photogenerated carriers, thus reducing both the electron–hole recombination and the photoresponse time. Since GQDs have a good response to UV light, sensitization with GQDs may, on the other hand, enhance the overall UV absorption in ZnO and hence the photoconductivity of ZnO-based photodetector. The pronounced charge separation at the GQDs/ZnO interface also provides good prerequisites of using this hybrid material in perovskite-, dye sensitized-, quantum-dot-sensitized- and inverted polymer solar cells[30–43]. ZnO–graphene quantum dots system is promising for designing optoelectronic devices (namely light-emitting diodes, LEDs)[44–49], playing a role of either electron transport layer or active emissive component. Particularly, a brightness of 798 cd·m$^{-2}$ was achieved for white LEDs based on ZnO cores wrapped in a shell of GQDs[49].

One more important application of ZnO–GQDs hybrid nanomaterials is related to their excellent photocatalytic properties. The enhanced light absorption in such combined systems and, consequently, increased number of photo-induced charge carriers make it possible to reach synthesis of tetrasubstituted propargylamines[50], improved solar-driven water splitting[51] and photocatalytic $H_2$ evolution under visible light[52], selective reduction of nitroarenes[53], and effective degradation of metronidazole (MNZ) antibiotic[54], colored pollutants (Rhodamine B, methylene blue (MB) and methylene orange (MO))[55–61], colorless pollutant (carbendazim (CZ) fungicide)[62] and glyphosate herbicide contaminated in agricultural wastewater[63].

Effective interfacial charge transfer between GQDs and ZnO under photoirradiation also facilitates the production of reactive oxygen species (ROS), which make it possible to use this material as an antibacterial agent[64–68]. Indeed, the formation of ROS may activate the electrostatic attraction between ZnO–GQDs hybrid and target bacteria (*Escherichia coli, Pseudomonas aeruginosa, Bacillus cereus and Staphylococcus aureus*), thereby causing a damage of bacterial cell membrane and inhibiting the growth of bacterial colonies.

All mentioned here examples highlight the importance of a deep understanding of the fundamental nature of photoexcitation and charge transport in ZnO–GQDs hybrid materials. Reaching such an understanding will provide guidelines for application-specific designs of ZnO-GQDs nanohybrids. Although the fundamentals of the response of ZnO–GQDs to light upon photoexcitation have been described in detail in literature, the main focus of the existing investigations is on a physical combination of ZnO and GQDs, each of which remains its own functional characteristics. In this case, both materials interact in a way with each other, but no formation of a new substance is expected. But what if we combine two isolated components into one stable material platform? Intuitively, such an approach could be conducive to avoid poor inter-component adhesion, thereby addressing overall temporal stability problem. Inspired by an earlier work of Quang *et al*. which is dedicated to the formation of graphene-like monolayer ZnO membranes suspended in graphene pores[69], here we propose a new concept of organic-inorganic solid-state hybridization based on atomically thin graphene-like ZnO clusters embedded into GQDs. This can be referred to a monolithic integration between ZnO and GQDs. Since such a material system is still unexplored, we employ density-functional theory (DFT) and time-dependent DFT (TD-DFT) calculations to explore the nature of photoexcited states in $(ZnO)_nC_{96-2n}$ system with $n$=1, 3, 4, 7, 12 and 27.

**Methodology**

All quantum chemistry calculations were conducted by means of Gaussian 16 Rev. C.01 program package[70] under gas-phase conditions. Circumcircumcoronene molecule ($C_{96}H_{24}$) demonstrated in **Figure S1** (**ESI**) was chosen as a representative of GQDs and a host organic molecule for ZnO clusters. ZnO-embedded graphene quantum dots, hereinafter referred to as $(ZnO)_nC_{96-2n}$ systems, were constructed through replacing carbon (C) atoms with zinc (Zn) and oxygen (O) atoms (see **Figure S2**, **ESI**) in a way that ensures balanced substitutional doping of the system of hexagonally arranged carbon atoms in $C_{96}H_{24}$. This is analogous to a symmetric doping of two sublattices of infinite graphene model. As a result, there were equal numbers of zinc and oxygen atoms, indicating that the stoichiometry of ZnO clusters embedded in graphene quantum dots is 1:1. Geometry optimization of the $(ZnO)_nC_{96-2n}$ systems was performed using Perdew-Burke-Ernzerhof exchange–correlation hybrid functional (PBE0, also called PBE1PBE)[71], which combines 25% exact Hartree-Fock (HF) exchange with 75% of PBE exchange. PBE0 is a reasonable choice to correctly reproduce the experimentally observed electronic and optical properties of ZnO[72] and GQDs[73,74,75]. 6-31G* basis sets[76] were used for carbon, hydrogen, and oxygen atoms, while the SDD basis set developed by the Stuttgart-Dresden-Bonn group[77] was used for zinc atom. $(ZnO)_nC_{96-2n}$ structures corresponding to the lowest energy minima were obtained by enabling a full relaxation of all coordinates using the default convergence criteria. Vibrational mode calculations were additionally carried out to make sure there are no imaginary frequencies and to confirm the equilibrium of the ground-state geometry. Finally, absorption spectra of the $(ZnO)_nC_{96-2n}$ hybrid materials were predicted by using the time-dependent density functional theory (TD-DFT) approach at the same level of theory (PBE1PBE/6-31G*, SDD). The lowest 100 excited states were considered. The analysis of excited states was performed by using Multiwfn program[78]. The VESTA program[79] was employed to visualize the structures of $(ZnO)_nC_{96-2n}$. The stability of $(ZnO)_nC_{96-2n}$ structures was tested by estimating the cohesive energy per atom[80]:

$$E_{coh} = \frac{E_{(ZnO)_nC_{96-2n}} - (\sum_i E_C + \sum_j E_H + \sum_k E_{Zn} + \sum_l E_O)}{n_C + n_H + n_{Zn} + n_O} \quad (1)$$

where $E_{(ZnO)_nC_{96-2n}}$ is the total energy of the relaxed $(ZnO)_nC_{96-2n}$ structure, $E_{C,H,Zn,O}$ is the energy of isolated carbon, hydrogen, zinc, and oxygen atoms, respectively. The summation indices $i$, $j$, $k$ and $l$ correspond to the total amount of carbon, hydrogen, zinc, and oxygen atoms ($n_{C,H,Zn,O}$) in $(ZnO)_nC_{96-2n}$ system with $n$=1, 3, 4, 7, 12 and 27.

## Results and discussion

The optimized geometric structures of $(ZnO)_nC_{96-2n}$ are illustrated in **Figure 1**, while the cohesive energies and selected structural parameters are reported in **Table 1**. First, it should be noted that all constructed hybrid structures are energetically stable (with quite high energy barriers for the decomposition of molecule into individual atoms), as evidenced by the negative cohesive energies. Such a predicted energetic stability implies the principal possibility of experimental design of $(ZnO)_nC_{96-2n}$ quantum dots. However, the cohesive energy tends to become less negative with increasing the number of ZnO pairs embedded into the $C_{96}$ matrix that entails a gradual decrease in the stability. The accommodation of ZnO pairs in the GQDs is achieved not only through the formation of new bonds like Zn-O, Zn-C and C-O, but also through a pronounced curvature of initially flat GQDs (**Figure S3**, **ESI**). The curvature parameter is estimated as the difference between the average $z$-coordinate and the absolute highest $z$-coordinate in the $(ZnO)_nC_{96-2n}$. The results are listed in **Table 1**. The difference between minimum and maximum $z$-coordinates is included in square brackets as an additional curvature parameter (**Table 1**). It is apparent that both curvature parameters demonstrate a significant increase with increasing the amount of incorporated ZnO pairs, following the trend of decreasing stability. In the limit case (when 27 ZnO molecules are incorporated into $C_{96}$ skeleton), i.e. $(ZnO)_{27}C_{42}$ system, to a large extend, resembles the hemisphere fullerene[81]. Interestingly, the deviations from an ideally planar atomic arrangement decrease Zn–O, C–C and C–O bonds and, concomitantly, cause slight Zn–C bond length elongation. The stability of $(ZnO)_nC_{96-2n}$ systems has been further confirmed by performing frequency calculations. The absence of the appreciable imaginary frequencies suggests the good stability of $(ZnO)_nC_{96-2n}$ hybrid materials. The most interesting part of predicted Raman spectra of the $(ZnO)_nC_{96-2n}$ that is related to the vibrations of light C atoms is demonstrated in **Figure 2a**. It is clearly seen that the incorporation of ZnO into GQDs (up to $n=12$) causes a reduction of the molecular symmetry that is manifested by the appearance of a large set of local vibrational modes instead of few Raman modes of pristine GQDs (including the most intensive one at 1374 cm$^{-1}$) within the spectral range from 1050 to 1650 cm$^{-1}$. It is striking to note that the $(ZnO)_{27}C_{42}$ structure has much stronger Raman activity than other considered structures including the reference pristine GQDs. The corresponding Raman spectrum is dominated by two bands at 1119 and 1384 cm$^{-1}$, which are assigned to the atomic movements of carbon at the edges of $(ZnO)_{27}C_{42}$ structure (around the entire perimeter). The observation of these intense Raman modes can be attributed to both plasmon resonance effects and an improvement of molecular symmetry. The latter is further confirmed by the molecular orbital analysis, according to which the higher occupied molecular orbital (HOMO) and second unoccupied orbital are doubly degenerate. The incorporation of small amount of ZnO molecules reduces degeneracy of orbitals and causes an energy-level splitting, as was observed for the structures with $n$ from 1 to 7. However, the fact that the lowest energy orbitals for both $(ZnO)_{12}C_{72}$ and $(ZnO)_{27}C_{42}$ systems remain degenerate with each other, like in the case of unperturbed $C_{96}$ system, highlights the symmetric configuration of these two hybrid molecules. Much stronger Raman activity of $(ZnO)_{27}C_{42}$ compared to that of $(ZnO)_{12}C_{72}$ can be explained by the abundance of conduction electrons participating in collective oscillations of free charges and hence contributing to the plasmon resonance.

From **Table 1**, it is also seen that the HOMO-LUMO energy gap decreases upon increasing the amount of ZnO molecules embedded into $C_{96}$ matrix. The minimum energy gap of ~1.24 eV is achieved for $(ZnO)_{27}C_{42}$ system (**Figure S4, ESI**). The observed energy gap narrowing is mainly affected by the upward shift of the HOMO level, while the downward shift of LUMO acquires a less pronounced character.

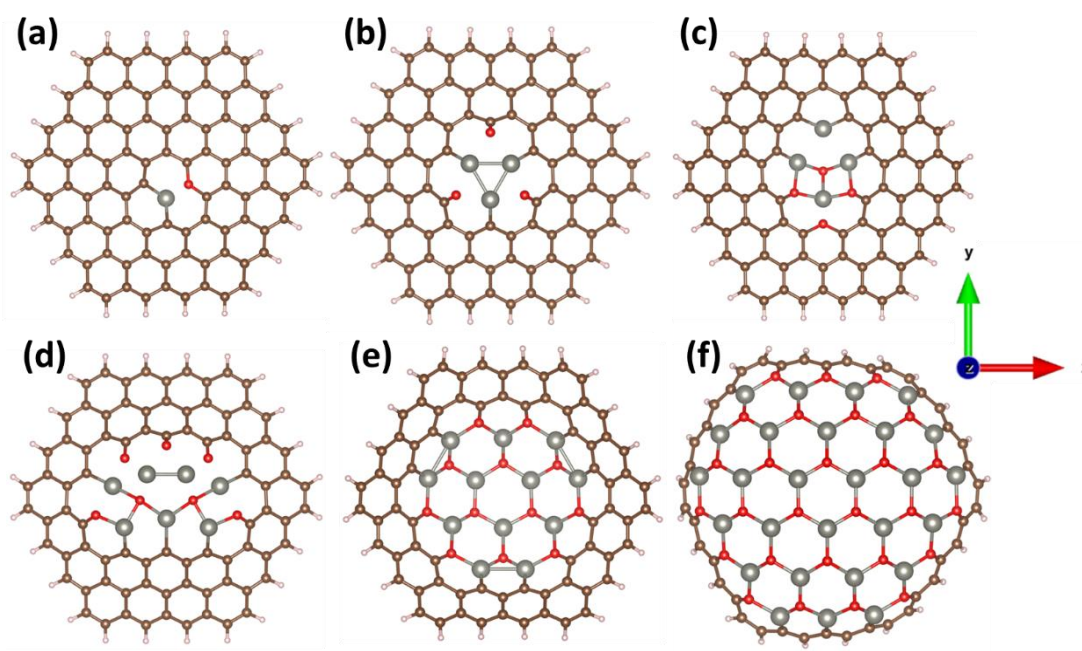

**Figure 1**. (Top view) Optimized structures of the $(ZnO)_nC_{96-2n}$: (a) $n=1$, (b) $n=3$, (c) $n=4$, (d) $n=7$, (e) $n=12$ and (f) $n=27$, respectively. Brown, whitish, blue, and red balls correspond to carbon, hydrogen, zinc, and oxygen atoms, respectively.

**Table 1.** Cohesive energies, structural parameters, and energy gap values of $(ZnO)_nC_{96-2n}$ hybrid structures.

| Number of ZnO pairs, $n$ | Cohesive energy, eV | Mean Zn-O bond length, Å | Mean C-C bond length, Å | Mean C-O bond length, Å | Mean C-Zn bond length, Å | Curvature parameter, Å | Energy gap, eV |
|---|---|---|---|---|---|---|---|
| 0 | -8.2801 | - | 1.4302 | - | - | 0.0053 [0.0106] | 3.1184 |
| 1 | -8.1190 | 2.0864 | 1.4167 | 1.3971 | 1.8748 | 2.0937 [3.1488] | 2.5981 |
| 3 | -7.9128 | 2.1315 | 1.4197 | 1.2828 | 1.9416 | 2.4904 [3.4627] | 1.9519 |
| 4 | -7.8183 | 1.9652 | 1.4178 | 1.3744 | 1.8949 | 3.4944 [5.2536] | 1.9731 |
| 7 | -7.5421 | 2.0383 | 1.4198 | 1.3118 | 1.9439 | 3.4653 [5.1961] | 1.3331 |
| 12 | -7.1017 | 1.9043 | 1.4211 | 1.3681 | 1.8909 | 3.8044 [5.8253] | 1.7992 |
| 27 | -5.8035 | 1.9039 | 1.4091 | 1.3710 | 1.9396 | 3.7695 [6.8327] | 1.2436 |

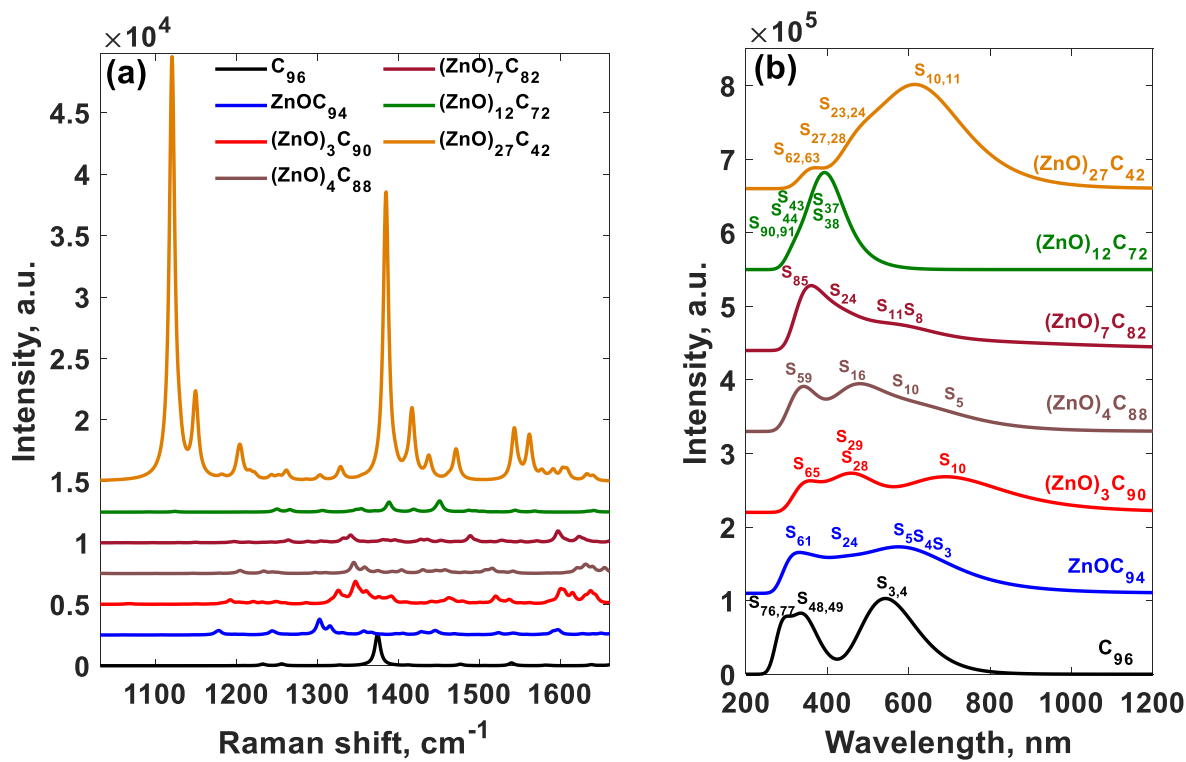

**Figure 2.** Predicted Raman spectra (a) and absorption spectra (b) of the $(ZnO)_nC_{96-2n}$ structures, respectively.

The results of electron excitation analysis also speak in favour of the supposition that the symmetry of the hybrid molecule affects its electronic properties. From **Figure 2b** is it apparent that only highly symmetric molecules ($C_{96}$, $(ZnO)_{12}C_{72}$ and $(ZnO)_{27}C_{42}$) exhibit doubly degenerate excited state configurations. The dominant spectral features observed at 290.9, 335.4, and 542.8 nm in the $C_{96}$ spectrum are assigned to $S_{76,77} \leftarrow S_0$, $S_{48,49} \leftarrow S_0$, and $S_{3,4} \leftarrow S_0$ transitions, respectively. Considering the electron–hole wave-function overlap integral ($S_r$), the distance between centroids of holes and electrons (*D*) and the degree of separation of holes and electrons (*t*) as descriptors of type of electron excitation[82,83,84], it is possible to draw some conclusions about the nature of these transitions. From **Table 2** (see also **Table S1** summarizing the transitions with oscillator strength >0.1), it is clear that *D* of all aforementioned excited states is close to zero, while the *S* parameters are 0.78, 0.87 and 0.92 for $S_{76,77} \leftarrow S_0$, $S_{48,49} \leftarrow S_0$, and $S_{3,4} \leftarrow S_0$, respectively. *t* indices are negative and are much less than 0, pointing out the absence of the charge separation. This suggests that the corresponding excitations are typical locally excited (LE) states. This finding is then corroborated by the analysis of the charge density difference (CDD) between ground state and excited state (**Figure S5**, **ESI**). Particularly, it was revealed that hole and electrons are delocalized over the entire area of pristine $C_{96}$ system, and no charge separation occurs. **Table 2** also summarises the electronic transitions that are involved in these excited states.

The absorption spectrum of the ZnO-free pristine molecule undergoes substantial changes after ZnO incorporation (**Figure 2b**, **Table 2**). A quick look at the evolution of the absorption spectra shows that the symmetry breaking significantly red shifts the main absorption bands of $C_{96}$ and reduces the absorption intensity (the oscillator strength for the corresponding transitions), as can be clearly seen in the case of low-symmetry intermediate systems (from $ZnOC_{94}$ to $(ZnO)_7C_{82}$). However, a further increase of inserted ZnO molecules leads to a distinct increase of the oscillator strength and hence absorption intensity. Finally, highly symmetric $(ZnO)_{27}C_{42}$ molecule demonstrates a very strong absorbance spectrum extending from ultraviolet to near infrared, with oscillator strength even higher than that for the localized excitation in the case of pristine $C_{96}$ system.

A more detailed consideration of each of the presented spectra shows that in the case of $ZnOC_{94}$ system the most intense band extending from 400 nm to 900 nm is related to $S_3$-$S_5$ excited states, respectively. The complementary analysis of the CDD (**Figure 3a**), $S_r$, $D$ and $t$ parameters (**Table 2**) shows that all three excitations are LE–CT hybrid states. Interestingly, the electrons and holes are more separated in $S_3$ and $S_5$ states than in $S_4$ state (that is confirmed by larger $D$ and less negative $t$), indicating a dominant CT character for these excited states. For $(ZnO)_3C_{90}$ system, the distance between centroids of holes and electrons for three most probable transitions is very low because they occupy the same space (**Figure 3b**). Therefore, $S_{10}$, $S_{28}$ and $S_{29}$ states are mostly local excitations. $(ZnO)_4C_{88}$ exhibits a relatively complex series of spectral features, with dominant contribution from $S_5$, $S_{10}$ and $S_{16}$ excited states. First two states can be identified as LE–CT hybrid states with prevailing CT character (**Figure 3c**), while the $S_{16}$ state is characterized by low $D$ and rather negative $t$ values, which are peculiar to LE state with minimal charge separation. The absorption spectrum $(ZnO)_7C_{82}$ is dominated by a wide band peaked at 365 nm (with major contribution from $S_{85} \leftarrow S_0$ transition) followed by the weak shoulder at ~612 nm that can be attributed to the $S_8 \leftarrow S_0$, and $S_{11} \leftarrow S_0$ transitions. Unexpectedly, all three transitions have completely different nature (see **Table 2** and **Figure 3d**). $S_8$ and $S_{85}$ can be regarded as local excitations, while $S_{11} \leftarrow S_0$ has obvious charge transfer characteristics. Furthermore, it is important to note that among all considered systems only $(ZnO)_7C_{82}$ exhibits the excitation with the largest $D$ parameter (6.29 Å) and positive $t$ index (1.79 Å). The sharp spectral feature of the $(ZnO)_{12}C_{72}$ molecule at ~390 nm is preferentially attributed to three doubly degenerate LE–CT hybrid states: $S_{37,38}$, $S_{52,53}$ and $S_{43,44}$, respectively (**Table 2**, **Figure 5e; see also Figure S6a, ESI**).

**Table 2**. Properties of dominant excited states in $(ZnO)_nC_{96-2n}$ hybrid systems. For doubly degenerate excite states, the properties of second sate adjacent to first state are included in square brackets.

| System | Excited state | Wavelength, nm | Oscillator strength, $f$ | $D$, Å | $S_r$ | $t$, Å | Type | Major contribution |
|---|---|---|---|---|---|---|---|---|
| $C_{96}$ | $S_{3,4}$ | 542.86 | 1.27 | 0 | 0.96 | -1.94 | LE | H-1->LUMO (43%), HOMO->L+1(43%) [H-1->L+1 (43%), HOMO->LUMO (43%)] |
| | $S_{48,49}$ | 335.47 | 0.45 | 0 | 0.95 | -2.61 | LE | H-8->LUMO (11%), H-7->L+1(11%), H-4->L+3 (22%), H-3->L+4 (22%) [H-8->L+1(11%), H-7->LUMO (11%), H-4->L+4 (22%), H-3->L+3 (22%)] |
| | $S_{76,77}$ | 290.95 | 0.46 | 0 | 0.91 | -3.81 | LE | H-5->L+3 (58%) [H-5->L+4 (58%)] |
| $ZnOC_{94}$ | $S_3$ | 675.22 | 0.28 | 1.97 | 0.84 | -1.97 | LE-CT | H-2->LUMO (18%), H-1->LUMO (26%), HOMO->L+1 (38%) |
| | $S_4$ | 628.63 | 0.26 | 1.06 | 0.85 | -2.82 | LE-CT | H-1->L+1 (49%), HOMO->L+2 (34%) |
| | $S_5$ | 605.3 | 0.26 | 2.03 | 0.83 | -1.98 | LE-CT | H-1->L+1 (21%), HOMO->L+2 (55%) |
| $(ZnO)_3C_{90}$ | $S_{10}$ | 696.97 | 0.71 | 0.52 | 0.86 | -3.63 | LE | H-4->LUMO (49%), H-3->LUMO (13%), H-1->L+2 (10%), HOMO->L+1 (21%) |
| | $S_{28}$ | 474.2 | 0.27 | 0.47 | 0.89 | -4.16 | LE | H-4->L+1 (20%), H-2->L+2 (46%) |
| | $S_{29}$ | 472.66 | 0.32 | 0.69 | 0.86 | -3.58 | LE | H-12->LUMO (51%), H-2->L+1 (11%), H-1->L+6 (12%) |
| $(ZnO)_4C_{88}$ | $S_5$ | 658.09 | 0.41 | 3.78 | 0.80 | -0.53 | LE-CT | H-1->LUMO (17%), H-1->L+2 (18%), HOMO->L+1 (57%) |
| | $S_{10}$ | 532.56 | 0.44 | 1.67 | 0.91 | -2.95 | LE-CT | H-4->LUMO (10%), H-2->L+1 (54%), HOMO->L+2 (16%) |
| | $S_{16}$ | 480.26 | 0.36 | 1.47 | 0.87 | -3.05 | LE | H-5->LUMO (11%), H-3->LUMO (31%), H-1->L+4 (26%), H-1->L+5 (21%) |
| $(ZnO)_7C_{82}$ | $S_8$ | 612.42 | 0.39 | 0.61 | 0.75 | -1.57 | LE | HOMO->L+1 (84%) |
| | $S_{11}$ | 555.71 | 0.27 | 5.97 | 0.65 | 1.58 | CT | HOMO->L+2 (85%) |
| | $S_{85}$ | 341.34 | 0.28 | 0.52 | 0.85 | -3.36 | LE | H-5->L+7 (15%), H-4->L+8 (38%) |
| $(ZnO)_{12}C_{72}$ | $S_{37,38}$ | 407.09 | 0.60 | 1.53 | 0.81 | -2.47 | LE-CT | H-9->LUMO (16%), H-4->L+3 (18%), H-3->L+6 (13%), H-2->L+5 (13%) [H-8->LUMO (16%), H-4->L+4 (18%), H-3->L+5 (13%), H-2->L+6 (13%)] |
| | $S_{43,44}$ | 388.26 | 0.38 | 1.30 | 0.85 | -2.69 | LE-CT | H-11->LUMO (12%), H-4->L+3 (16%), HOMO->L+7 (38%) [H-10->LUMO (12%), H-4->L+4 (16%), H-1->L+7 (38%)] |
| | $S_{52,53}$ | 375.48 | 0.21 | 2.35 | 0.82 | -2.68 | LE-CT | H-4->L+6 (13%), HOMO->L+8 (61%) [H-4->L+5 (13%), H-1->L+8 (61%)] |
| $(ZnO)_{27}C_{42}$ | $S_{10,11}$ | 635.52 | 1.55 | 0.24 | 0.89 | -5.25 | LE | H-1->L+4 (46%), HOMO->L+3 (46%) [H-1->L+3 (46%), HOMO->L+4 (46%)] |
| | $S_{23,24}$ | 496.79 | 0.51 | 1.06 | 0.88 | -4.41 | LE-CT | H-2->L+4 (15%), H-1->L+9 (11%), HOMO->L+8 (40%), HOMO->L+9 (19%) [H-2->L+3 (15%), H-1->L+8 (40%), H-1->L+9 (19%), HOMO->L+9 (11%)] |
| | $S_{27,28}$ | 466.6 | 0.27 | 1.33 | 0.86 | -3.63 | LE-CT | H-3->L+3 (19%), H-3->L+4 (10%), H-1->L+8 (12%), HOMO->L+9 (32%) [H-3->L+3 (10%), H-3->L+4 (19%), H-1->L+9 (32%), HOMO->L+8 (12%)] |

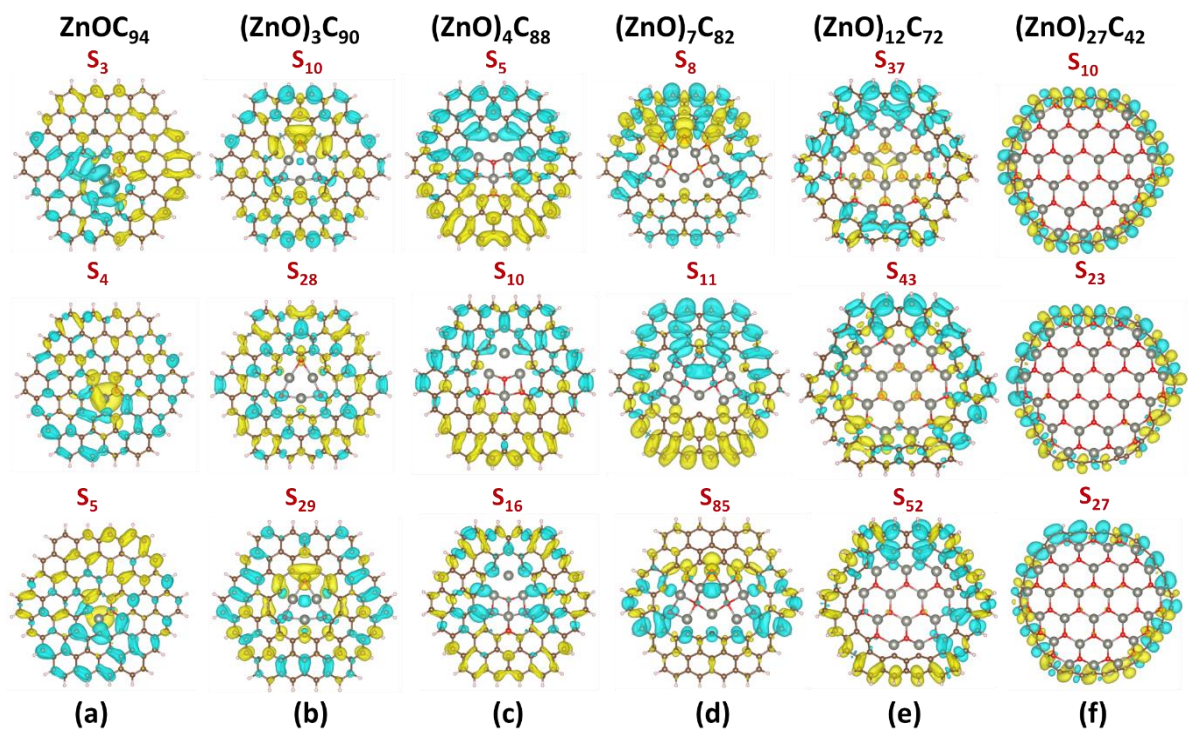

**Figure 3.** CDD for selected excited states (with the largest oscillator strength) in (a) $ZnOC_{94}$, (b) $(ZnO)_3C_{90}$, (c) $(ZnO)_4C_{88}$, (d) $(ZnO)_7C_{82}$, (e) $(ZnO)_{12}C_{72}$, and (f) $(ZnO)_{27}C_{42}$, respectively. Herein, yellow denotes positive charge distribution and cyan means negative charge distribution. CDD was calculated as a difference between the corresponding excited state and the ground state of the considered system ($\rho_{\text{excited}} - \rho_{\text{ground}}$). Isosurface level is set to be 0.0003.

Apparently, for $(ZnO)_{27}C_{42}$, the doubly degenerate $S_{10,11}$ excited state possessing the largest oscillator strength of 1.55 is a typical LE state, while two minor doubly degenerate states ($S_{23,24}$ and $S_{27,28}$) demonstrate both LE and CT characteristics (**Table 2**, **Figure 5f; see also Figure S6b, ESI**). The nature of the wide absorption band observed at 614 nm is further scrutinized through the decomposition of the absorption spectrum into four different components. This method originally developed by Liu *et al.*[85] enables distinguishing between intrafragment charge redistribution and interfragment charge transfer. In other words, by performing such an analysis it is possible to conclude about the contribution of the charge-transfer excited states to the total response of the system. In line with this we selected two different fragments in $(ZnO)_{27}C_{42}$ – $C_{42}$ and $(ZnO)_{27}$ – and then calculated the so-called charge-transfer spectra (CTS) of $(ZnO)_{27}C_{42}$ molecule (**Figure 4a**). Even though the charge-transfer components ($C_{42} \rightarrow (ZnO)_{27}$ and $(ZnO)_{27} \rightarrow C_{42}$, respectively) contribute to the total absorption spectra, it is seen that the band at 614 nm is mostly originating from the electron transitions within the $C_{42}$ moiety, additionally pointing to the domineering role of local excitations in this spectral region. Therefore, it can be stated that zinc and oxygen atoms are minimally involved in the corresponding excitations since the electrons and holes are mostly localized at the peripheral carbon moieties. In contrast to this, the analysis of the weak absorption band at 366 nm that is dominated by doubly degenerate $S_{62,63}$ states (**Figure 2b**), demonstrates a significant role of the charge transfer from the $C_{42}$ moiety to the $(ZnO)_{27}$ part in the $S_{62,63}$ excitations and a small contribution of electronic transitions within the $(ZnO)_{27}$ fragment. CDD analysis provides additional evidence of the origin of $S_{62,63}$ states (**Figure 4b**). Particularly, we notice that the electron accumulation region is mainly located at the $(ZnO)_{27}$ fragment, while the obvious charge depletion region is observed at $C_{42}$ moiety.

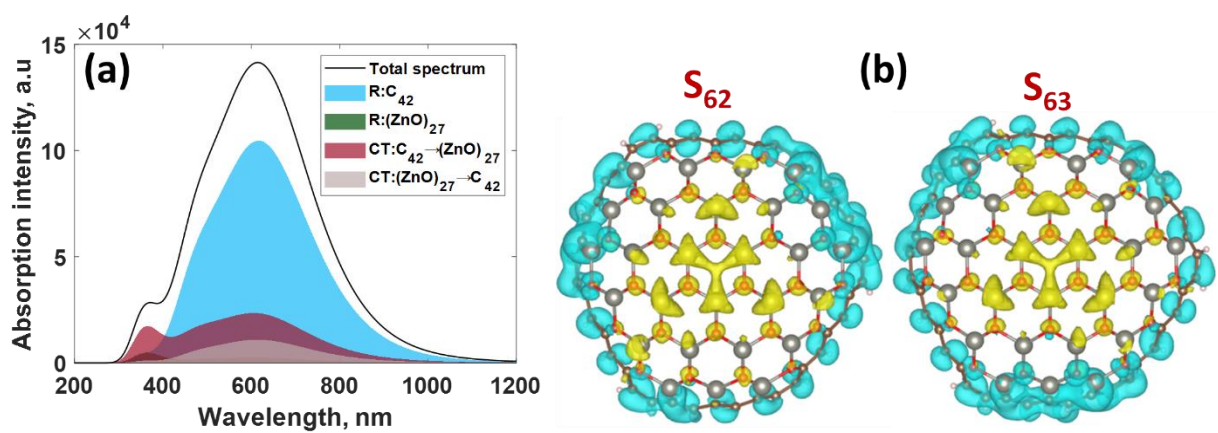

**Figure 4.** (a) Charge-transfer spectrum of the $(ZnO)_{27}C_{42}$ including intrafragment electron redistribution (R) components and interfragment charge transfer (CT) contributions. (b) CDD for $S_{62,63}$ excited states in $(ZnO)_{27}C_{42}$. Yellow corresponds to positive charge distribution and cyan denotes negative charge distribution. Isosurface level is set to be 0.0002.

## Conclusions

By performing geometry optimization and frequency calculations using PBE0/6-31G*/SDD level of DFT, stable $(ZnO)_nC_{96-2n}$ hybrid systems with $n$=1, 3, 4, 7, 12 and 27 were designed. The increase of amount of ZnO pairs embedded in $C_{96}$ matrix caused the reduction of the cohesive energy of the system, which nonetheless remains negative, indicating a principal possibility to synthesize these hybrid materials. By regulating ZnO content one can possible alter the symmetry and curvature of the ZnO-embedded graphene quantum dots, HOMO-LUMO energy gap, and hence light absorption properties including the dominant absorption wavelength, excitation type and oscillator strength. Highly symmetric $(ZnO)_{27}C_{42}$ molecule was found to demonstrate the enhanced Raman signal mostly originating from the collective vibrations within the peripheral $C_{42}$ moiety and a strong broad absorption band extending from UV to near infrared. These make such material promising for designing both SERS-based sensors and cost-effective broadband solar absorbers. The in-depth analysis of the prevailing excited states enabled to shed light on their nature and to ascribe the strong absorption band in $(ZnO)_{27}C_{42}$ to the doubly degenerate locally excited states. The current work provides rich insights into the nature of the photoexcited states in ZnO-graphene hybrid materials and guidelines for further experimental design of organic-inorganic frameworks with desired charge-transfer characteristics.

## Acknowledgements

I.S. acknowledges the support from Ångpanneföreningens Forskningsstiftelse (Grant 21–112). All DFT calculations were enabled by resources provided by the Swedish National Infrastructure for Computing (SNIC), partially funded by the Swedish Research Council through grant agreement no. 2018-05973.

# Electronic Supplementary Information

# Nature of photoexcited states in ZnO-embedded graphene quantum dots

Ivan Shtepliuk and Rositsa Yakimova

*Semiconductor Materials Division, Department of Physics, Chemistry and Biology-IFM, Linköping University, S-58183 Linköping, Sweden*

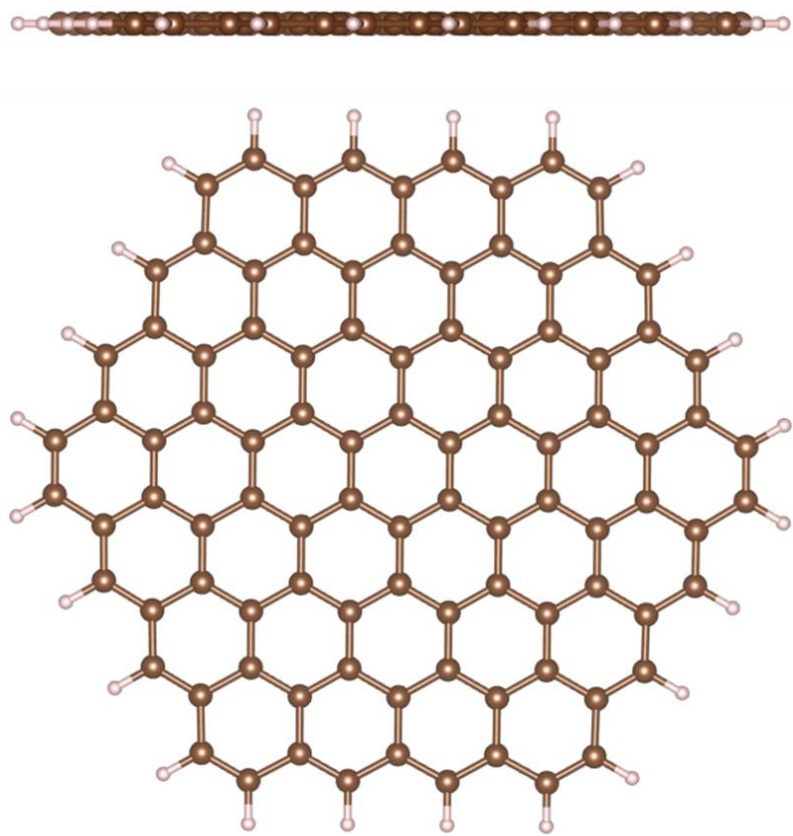

**Figure S1**. (Side and top views) Optimized structure of $C_{96}H_{24}$ molecule. Brown and whitish balls correspond to carbon and hydrogen atoms, respectively.

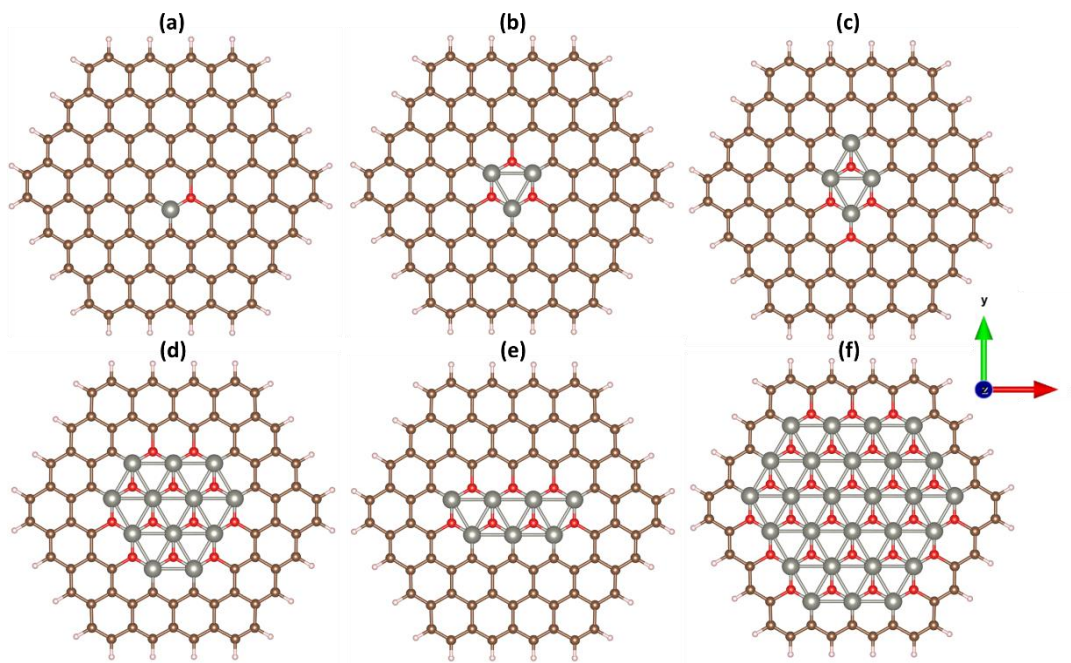

**Figure S2**. (Top view) Non-relaxed structures of the $(ZnO)_nC_{96-2n}$: (a) $n$=1, (b) $n$=3, (c) $n$=4, (d) $n$=7, (e) $n$=12 and (f) $n$=27, respectively. Brown, whitish, blue, and red balls correspond to carbon, hydrogen, zinc, and oxygen atoms, respectively.

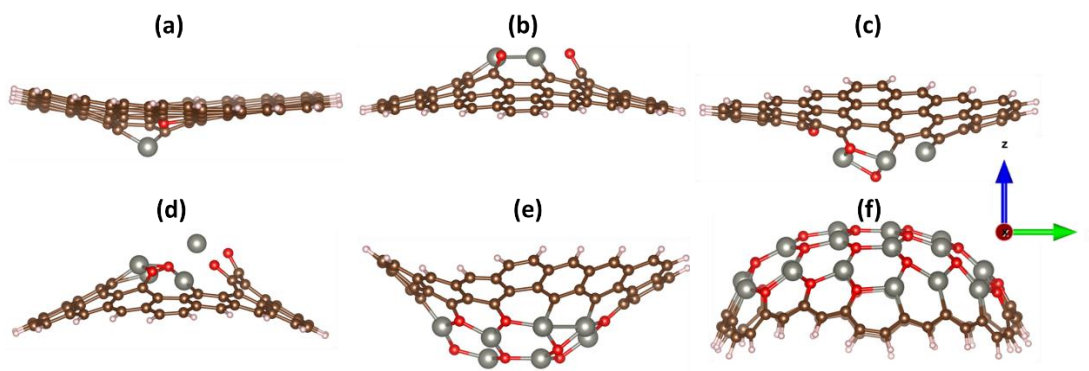

**Figure S3**. (Side view) Optimized structures of the $(ZnO)_nC_{96-2n}$: (a) $n$=1, (b) $n$=3, (c) $n$=4, (d) $n$=7, (e) $n$=12 and (f) $n$=27, respectively. Brown, whitish, blue, and red balls correspond to carbon, hydrogen, zinc, and oxygen atoms, respectively.

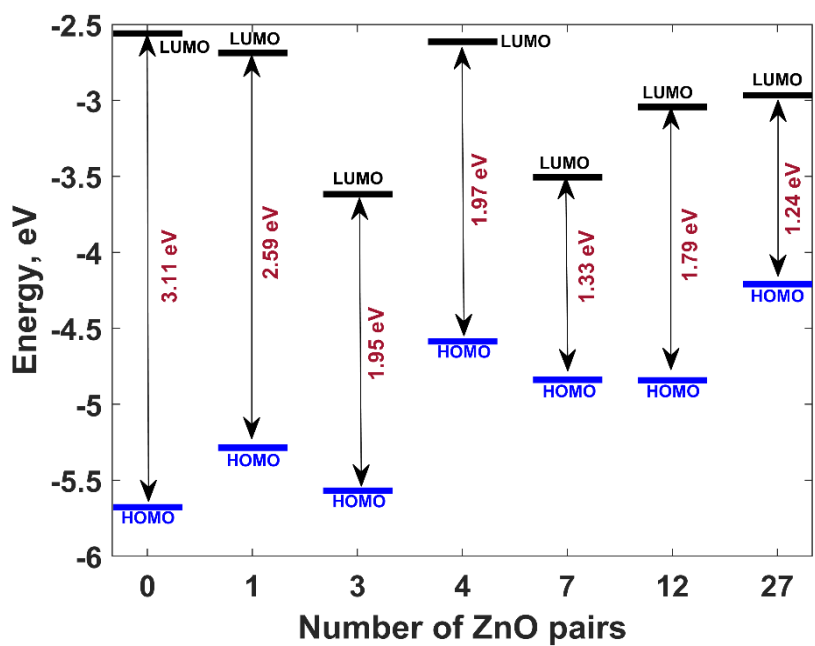

**Figure S4**. Molecular orbital diagram summarising all considered systems. The arrows designate HOMO–LUMO gaps.

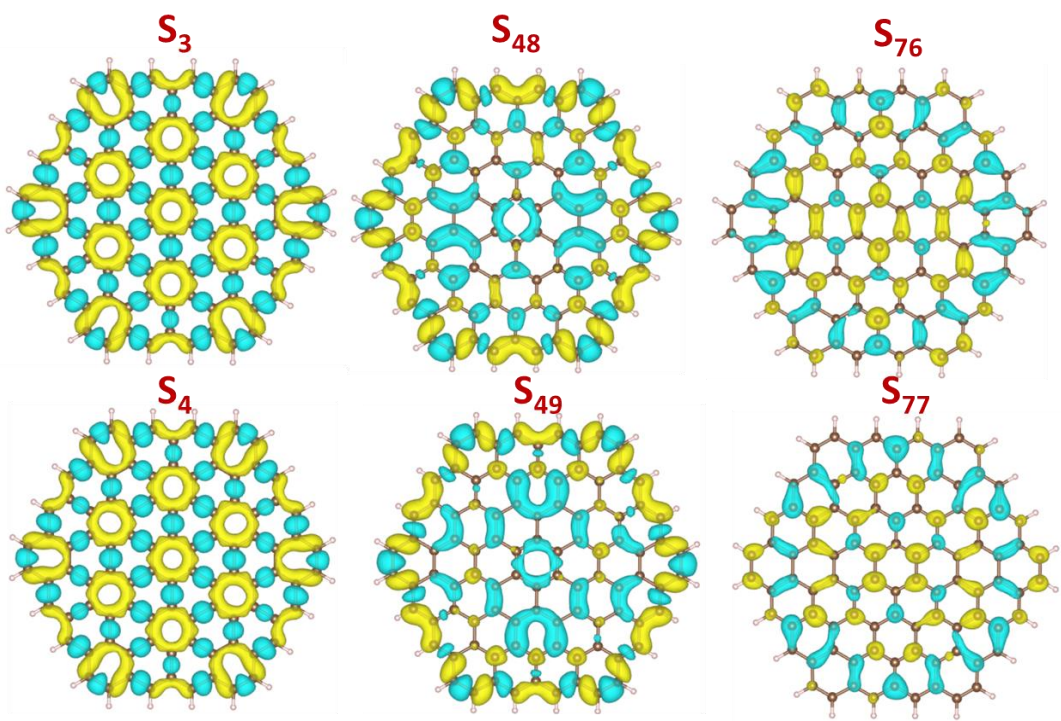

**Figure S5**. CDD for selected excited states (with the largest oscillator strength) in $C_{96}H_{24}$ molecule. Herein, yellow denotes positive charge distribution and cyan means negative charge distribution. Isosurface level is set to be $8\times10^{-5}$.

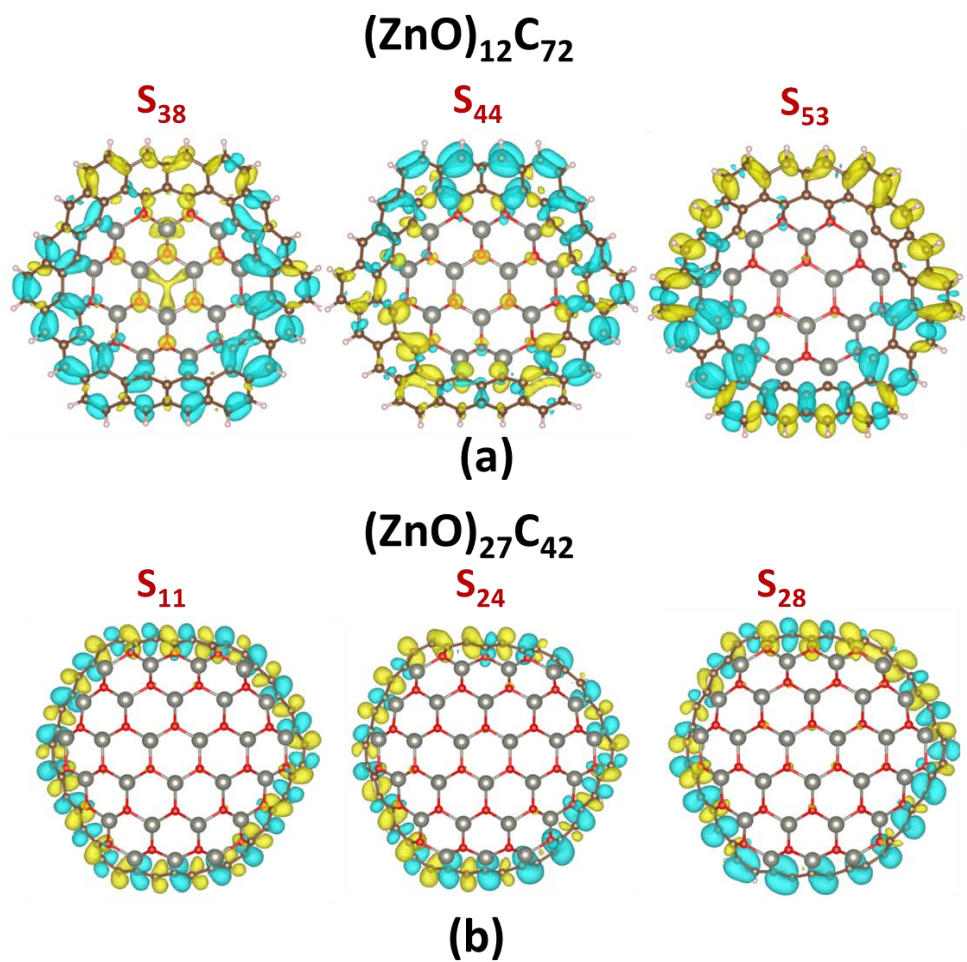

**Figure S6**. CDD for selected excited states (with the largest oscillator strength) in (a) $(ZnO)_{12}C_{72}$, and (b) $(ZnO)_{27}C_{42}$, respectively. Herein, yellow denotes positive charge distribution and cyan means negative charge distribution. Isosurface level is set to be 0.0003.

**Table S1.** Properties of excited states (with oscillator strength $f$ >0.1) in $(ZnO)_nC_{96-2n}$ hybrid systems. For doubly degenerate excite states, the properties of second sate adjacent to first state are included in square brackets.

| System | Excited state | Wavelength, nm | $f$ | $D$, Å | $S_r$ | $t$, Å | Type | Major contribution |
|---|---|---|---|---|---|---|---|---|
| $C_{96}$ | $S_{21, 22}$ | 381.22 | 0.16 | 0 | 0.83 | -3.92 | LE | H-6->LUMO (26%), H-3->L+2 (20%), H-1->L+5 (12%), H-1->L+8 (11%), HOMO->L+6 (12%) [H-6->L+1 (26%), H-4->L+2 (20%), H-1->L+6 (12%), HOMO->L+5 (12%), HOMO->L+8 (11%)] |
| | $S_{41, 42}$ | 345.27 | 0.1 | 0 | 0.88 | -2.87 | LE | H-6->LUMO (16%), H-3->L+2 (10%), H-2->L+3 (14%), HOMO->L+9 (50%) [H-6->L+1 (16%), H-4->L+2 (10%), H-2->L+4 (14%), H-1->L+9 (50%)] |
| | $S_{46, 47}$ | 341.27 | 0.24 | 0 | 0.80 | -1.26 | LE | H-9->LUMO (61%), H-4->L+4 (11%), H-3->L+3 (11%) [H-9->L+1 (61%), H-4->L+3 (11%), H-3->L+4 (11%)] |
| | $S_{72,73}$ | 296.42 | 0.24 | 0 | 0.77 | -4.50 | LE | H-13->L+1(21%),H-12->LUMO(21%),H-3->L+7 (42%) [H-13->LUMO(21%),H-12->L+1(21%),H-4->L+7 (42%)] |
| $ZnOC_{94}$ | $S_6$ | 590.45 | 0.23 | 1.75 | 0.73 | -2.50 | LE-CT | H-2->LUMO (58%), H-1->L+2 (13%) |
| | $S_7$ | 584.08 | 0.16 | 1.03 | 0.81 | -3.30 | LE-CT | H-2->L+1 (25%), H-1->L+2 (44%) |
| | $S_9$ | 543.52 | 0.20 | 3.36 | 0.67 | -0.86 | LE-CT | H-3->LUMO (62%) |
| | $S_{10}$ | 533.93 | 0.15 | 1.01 | 0.70 | -2.19 | LE-CT | H-2->L+1 (28%), H-1->L+2 (12%), H-1->L+3 (22%) |
| | $S_{13}$ | 494.15 | 0.14 | 2.33 | 0.70 | -1.91 | LE-CT | H-3->L+1 (41%), HOMO->L+5 (27%) |
| | $S_{17}$ | 467.15 | 0.12 | 0.47 | 0.77 | -3.24 | LE | H-4->LUMO(23%),H-2->L+2(16%),HOMO->L+4(19%) |
| | $S_{24}$ | 437.99 | 0.20 | 1.90 | 0.68 | -1.95 | LE-CT | H-3->L+2 (29%), H-2->L+2 (18%) |
| | $S_{44}$ | 369.50 | 0.15 | 1.41 | 0.63 | -3.03 | LE-CT | H-8->L+1(12%), H-3->L+4 (26%), HOMO->L+10 (13%) |
| | $S_{61}$ | 341.36 | 0.17 | 2.55 | 0.60 | -1.53 | LE-CT | H-5->L+3 (16%), H-3->L+6 (26%) |
| | $S_{100}$ | 298.22 | 0.12 | 0.28 | 0.63 | -4.01 | LE | H-15->L+1 (16%) |
| $(ZnO)_3C_{90}$ | $S_8$ | 747.47 | 0.19 | 0.56 | 0.83 | -2.16 | LE | H-7->LUMO (16%), HOMO->L+2 (69%) |
| | $S_{12}$ | 671.20 | 0.15 | 1.16 | 0.60 | -2.65 | LE-CT | H-8->LUMO(17%),H-7->LUMO(54%),H-5->LUMO (13%) |
| | $S_{30}$ | 466.26 | 0.12 | 0.52 | 0.77 | -3.64 | LE | H-4->L+1 (42%), H-3->L+1 (32%) |
| | $S_{35}$ | 450.57 | 0.10 | 0.39 | 0.81 | -4.71 | LE | H-4->L+2 (70%) |
| | $S_{65}$ | 356.62 | 0.18 | 0.87 | 0.64 | -1.99 | LE | H-3->L+4 (19%), H-2->L+5 (59%) |
| $(ZnO)_4C_{88}$ | $S_8$ | 603.21 | 0.14 | 0.70 | 0.82 | -3.14 | LE | H-2->LUMO (26%), H-1->L+2 (14%), H-1->L+3 (44%) |
| | $S_{15}$ | 492.31 | 0.10 | 1.84 | 0.62 | -0.31 | LE-CT | H-2->L+2 (38%), H-2->L+3 (16%), H-1->L+6 (20%), HOMO->L+4 (13%) |
| | $S_{20}$ | 464.44 | 0.10 | 0.82 | 0.73 | -2.86 | LE | H-2->L+3 (12%), H-1->L+6 (19%), H-1->L+7 (42%), HOMO->L+8 (10%) |
| | $S_{23}$ | 453.30 | 0.29 | 1.32 | 0.72 | -2.87 | LE-CT | H-5->LUMO (62%) |
| | $S_{32}$ | 420.62 | 0.10 | 0.10 | 0.72 | -1.22 | LE | H-5->L+1 (57%), HOMO->L+8 (16%) |
| | $S_{43}$ | 386.54 | 0.18 | 0.84 | 0.63 | -3.67 | LE | H-5->L+2 (61%) |
| | $S_{59}$ | 353.30 | 0.18 | 1.88 | 0.74 | -2.88 | LE-CT | H-7->L+2 (17%), H-4->L+4 (34%), H-3->L+5 (12%) |
| | $S_{60}$ | 349.29 | 0.15 | 0.44 | 0.73 | -0.91 | LE | H-7->L+2 (13%), H-2->L+9 (15%), H-1->L+13 (31%) |
| | $S_{65}$ | 343.38 | 0.15 | 1.97 | 0.66 | -2.57 | LE-CT | H-10->L+1 (46%), H-4->L+4 (12%), H-3->L+4 (13%) |
| | $S_{81}$ | 324.03 | 0.11 | 2.79 | 0.54 | -0.64 | LE-CT | H-6->L+5 (14%), H-4->L+7 (36%) |
| $(ZnO)_7C_{82}$ | $S_{24}$ | 464.04 | 0.23 | 4.57 | 0.61 | 0.37 | CT | H-9->LUMO (10%), H-3->L+2 (73%) |
| | $S_{30}$ | 437.17 | 0.10 | 2.63 | 0.59 | -0.90 | LE-CT | H-15->LUMO (28%), H-14->LUMO (20%), H-1->L+3 (10%), HOMO->L+6 (15%) |
| | $S_{31}$ | 435.36 | 0.12 | 0.83 | 0.62 | -2.60 | LE | H-1->L+3 (73%) |
| | $S_{48}$ | 392.34 | 0.13 | 3.17 | 0.64 | -0.52 | LE-CT | H-2->L+4 (23%), H-2->L+6 (15%), H-1->L+8 (28%) |
| | $S_{57}$ | 377.02 | 0.13 | 1.92 | 0.69 | -2.34 | LE-CT | H-4->L+4 (40%), HOMO->L+9 (11%) |
| | $S_{65}$ | 366.27 | 0.19 | 0.83 | 0.67 | -2.98 | LE | H-5->L+4 (30%), H-4->L+5 (15%) |
| | $S_{66}$ | 365.20 | 0.13 | 1.68 | 0.56 | -0.36 | LE-CT | H-5->L+3 (41%), H-4->L+6 (29%) |
| | $S_{88}$ | 338.03 | 0.11 | 0.53 | 0.69 | -3.32 | LE | H-10->L+1 (11%), H-8->L+2 (14%), H-5->L+7 (10%), H-3->L+9 (32%) |
| | $S_{93}$ | 334.17 | 0.12 | 0.45 | 0.64 | -3.04 | LE | H-5->L+7 (43%), H-2->L+10 (20%) |
| $(ZnO)_{12}C_{72}$ | $S_{32,33}$ | 414.94 | 0.12 | 2.56 | 0.42 | 0.52 | CT | H-10->LUMO (11%), H-8->LUMO (21%) [H-11->LUMO (11%), H-9->LUMO (21%)] |
| | $S_{61,62}$ | 362.51 | 0.20 | 1.94 | 0.66 | -1.91 | LE-CT | H-17->LUMO(15%), H-5->L+5 (23%), H-2->L+7 (38%) [H-16->LUMO (15%),H-5->L+6(23%),H-3->L+7 (38%)] |
| | $S_{90,91}$ | 323.30 | 0.16 | 2.34 | 0.53 | -1.68 | LE-CT | H-6->L+1 (32%), HOMO->L+16 (23%) [H-6->L+2 (32%), H-1->L+16 (23%)] |
| $(ZnO)_{27}C_{42}$ | $S_{19,20}$ | 535.45 | 0.10 | 1.52 | 0.79 | -3.86 | LE-CT | H-2->L+3 (68%) [H-2->L+4 (68%)] |